\documentclass[11pt,prd,superscriptaddress,nofootinbib]{revtex4}
\usepackage{graphicx}

\usepackage[usenames,dvipsnames]{xcolor}

\usepackage{amssymb,amsmath,amsfonts}
\title{sections and Chapters}

\begin{document}
	
	\hfill\parbox{5cm} { }
	
	\vspace{25mm}
	
	\begin{center}
		{	{\Large \bf  Meson-$\Delta$ and  meson-nucleon-$\Delta$  transition coupling constants in the soft-wall model of holographic QCD at finite temperature}
		
		\vskip 1. cm
		
		{Narmin Nasibova \footnote{Corresponding author: n.nesibli88@gmail.com}}}
		\vskip 0.5cm
	
		\it \indent  Institute of Physics, National Academy of Sciences,\\
		AZ-1148, H.Cavid ave. 131,  Baku, Azerbaijan\\
		\end{center}
	\thispagestyle{empty}
	\vskip2cm
	\centerline{\bf ABSTRACT}
	\vskip 4mm
In this paper,  minimal coupling constants of  meson-$\Delta$ baryon and meson-nucleon to $\Delta$ baryon transition have been investigated in the AdS/QCD soft-wall model at finite temperature limits. Initially, the right and left  profile functions of $\Delta$ baryons have been calculated using the Rarita-Schwinger   field in this model. After this, coupling constants have been written  by taking into account the profile functions of thermal hadrons  according to the zero-temperature case. Then, the temperature dependence graphs are plotted for strong couplings $g_{\rho \Delta \Delta}(T)$, $g_{a_{1} \Delta \Delta}(T)$ and  $g_{\rho N \Delta}(T)$. As a result, it has been  observed by an increase in  the temperature, values of the minimal coupling constants decreased and approached zero nearby Hawking temperature.

\section{Introduction}
One notable application of string theory is the anti-de Sitter/conformal field theory (AdS/CFT) \cite{1, 2, 3, 4, 5} or holographic duality, which is used to study the coupling regime of quantum field theories.
 This theory is also called gravity/gauge duality. This principle is based on closed and open string duality according to the relationship between the open and closed strings. As such, in this principle, the gravitational field corresponding to the closed string is defined by the boundary and the field theory corresponding to the open string by the bulk of the AdS space. QGP (Quark Gluon Plasma), condensed matter systems, and other QCD problems which is impossible to solve by helping perturbative theory have been investigated using this duality. The duality between gravitational and gauge theories is the strong-weak duality, which also was adapted to describe the low- and high-energy dynamics of Quantum Chromodynamics. It is also important to investigate hadron interaction processes in the nuclear medium in a proton-proton and heavy ion collisions in order to study phase transition of QGP and evolution of the early Universe.

 AdS/QCD (AdS/Quantum Chromodynamics) which was created based on this duality has two hard- and soft-wall models for modifying the metric in the AdS space-time. 
 These models are widely applied in particle physics for quantifying phenomenological prediction quantities and investigating other strong interaction problems.
 
In theoretical physics, many QCD problems were  solved by holographic QCD models in the vacuum and low-temperature medium in Refs. \cite{6, 7, 8, 9, 10, 11, 12, 13, 14}. It has been investigated the derivation of analytical formulas for the mass spectrum of mesons and $\Delta$ baryons and their coupling constants in Ref. \cite{15, 16}. In addition, it has been found reasonable results for the transition and electromagnetic form
factors of nucleons in the soft-wall model of holographic QCD at finite temperature.
Continuing these
investigations In Ref. \cite{17} we have studied the temperature dependence of vector meson coupling constant in the soft-wall model of AdS/QCD. It was observed the coupling constant of $\rho$ vector meson decreases on
temperature increases and vanishes at the temperature close to the confinement-deconfinement
phase transition temperature.

 It is interesting to check whether this situation takes place for
other meson sectors of the model as well.
As a continuation of Ref. \cite{17} we have decided to study, for the simplicity, minimal coupling constant of meson-baryon and meson-nucleon to $\Delta$ baryon transitions based on the case reviewed in Ref \cite{18} at finite temperature.
 Our aim here is to study temperature dependence in the spectrum of spin 3/2 baryons ($\Delta$ resonances) and the coupling constants between mesons and baryons  in the   soft-wall model of holographic QCD at finite temperature.
 
 The remainder of this paper is organized as follows:
Sections II and III are about the soft-wall model and the breaking of chiral symmetry at finite temperature.
In Sec. IV., the profile functions of $\Delta$ baryons have been obtained at finite temperature in the bulk of AdS space-time. 
In Sec.V, we construct the Lagrangian for the vector-spinor interaction in the bulk and obtain the temperature-dependent integral expression for the the minimal coupling constants of  $g_{\rho \Delta \Delta}(T)$, $g_{a_{1} \Delta \Delta}(T)$ and  $g_{\rho N \Delta}(T)$ using holographic correspondence at the boundary of QCD.
 In Sec. VI., the free parameters are fixed and the graphs of strong coupling constants are plotted at the different values of the quark flavor parameter $N_{f}$.
The last section is devoted to discussion and conclusions.

\section{Thermal SOFT-WALL MODEL of  holographic QCD}
The Schwarzschild metric can be written as follows \cite{15}:
\begin{eqnarray}
ds^{2}=e^{2A(z)}\left[f(z)dt^{2}-\left(d\vec{x}\right)^{2}-\frac{dz^2}{f(z)}\right],
\nonumber \\
 f(z)= 1-\frac{z^4}{z_H^4},
 \label{2}
\end{eqnarray}

 The relation between tortoise coordinate $r$ and $z$ is as follows:
\begin{equation*}
r=\int \frac{dz}{f(z)}
\end{equation*}
İn the finite temperature limit the r coordinate can be
expanded as
\begin{equation}
r\approx z\left[1+\frac{z^{4}}{5z_{H}^{4}}+\frac{z^{8}}{9z_{H}^{8}}\right].
\label{3}
\end{equation}
 So, AdS-Schwarzschild space time metric in the
Regge-Wheeler (RW) tortoise coordinate $r$ \cite{15} will be written as follows:
\begin{equation}
ds^{2}=e^{2A(r)}f^{\frac{3}{5}}(r)\left[dt^{2}-\frac{\left(d\vec{x}\right)^{2}}{f(r)}-dr^{2}\right]
\label{4}
\end{equation}
here $A(r)=log(\frac{R}{r})$ and the thermal factor $f(r)$ has the following form:
\begin{equation}
f(r)=1-\frac{r^{4}}{r_{H}^{4}}.
\label{5}
\end{equation}
  where, $r_{H}$ - is the position of the event horizon. 
 It is related to the Hawking temperature as $T=1/(\pi r_{H})$, $ x=(t,\vec x)$ is the set of Minkowski coordinates, $A(r)=log(\frac{R}{r})$, and  $R$ is the AdS space radius. 
  $k$ is a scale parameter.
   
    The AdS-Schwarzchild geometry is more suitable at high $T$, while at small $T$ this metric  can be also used by making a small $T$-expansion. The limit $T=0$ corresponds to a mapping of the AdS-Schwarzchild geometry onto AdS Poincare metric and the small $T$-behavior of hadron properties can be generated  in the formalism based on AdS Poincare metric and with the use of a thermal dilaton. It leads to equivalent results: AdS-Schwarzchild geometry with small $T$ is equal to AdS Poincare metric with thermal dilaton. 
   
   The dilaton field $\varphi$ has following form:
\begin{equation}
 \varphi(r,T)=K^{2}(T)r^{2},
 \label{6}
\end{equation}
where
\begin{equation}
K^{2}(T)=k^{2}\left[1+\rho(T) \right].
\label{7}
 \end{equation}
 So, $K^{2}(T)$ is the parameter in spontaneous breaking of chiral symmetry, while the thermal function $\rho (T)$ up to $T^4$ order has a form:
\begin{equation} 
\rho(T)=\delta_{T_{1}}\frac{T^{2}}{12F^{2}}+\delta_{T_{2}}\left(\frac{T^2}{12F^2}\right)^{2}.
\label{8}
\end{equation}
Here  $ \emph F$ is the decay constant and there are relation between the coefficients $\delta_{T_{1}}$ and  $\delta_{T_{2}}$ and  quark flavors $N_f$
  
\begin{equation}
\delta_{T_{1}}=-\frac{ N_{f}^{2}-1}{N_{f}},
\label{9}
 \end{equation}
 and
\begin{equation}
\delta_{T_{2}}=-\frac{N_{f}^{2}-1}{2N_{f}^{2}}. 
\label{10} 
\end{equation}
  
\section{CHIRAL symmetry breaking  at finite temperature}
In AdS/CFT correspondence, the currents in
the boundary  will correspond to the bulk gauge fields at finite temperature. The scalar field
 $X$ transforms as a bi-fundamental under the 
$SU(2)_L \times SU(2)_R$ group. The five-dimensional mass $ M^2_5 =
\Delta_{0}(\Delta_{0}-4)$ of the scalar which is fixed with the scaling dimension $\Delta_{0}$.
The solution for the bulk scalar $X$  at finite temperature is defined as,
$ \chi (r,T)\approx \frac{1}{2}am_{q}r+\frac{1}{2a}\Sigma (T)r^{3}=\vartheta (r,T) $ which the mass of quarks is $m_{q}$ and $\Sigma (T)$ is the value of thermal chiral condensate and $a=\frac{\sqrt N_{c}}{2\pi}$.
 The conventions in that the temperature dependence of $\Sigma(T)$ quark condensate is defined as $\Sigma(T)=\Sigma[1+\Delta_{T}]$ and 
$\Sigma(T)=< 0|\bar{q}q|0>_{T} = - N_{f}B(T)F^{2}(T)$.  In the chiral limit, $N_{f}$ is the number of quark flavors, $ B(T)$ is the quark condensate parameter, and $ F(T)$ is the pseudo-scalar meson decay constant in the chiral limit at finite temperature. $ F(T)$ and  $B(T)$ has been studied and calculated in \cite{19}.
\section{Meson Fields at finite temperature}

In this section, the author derives the results for the couplings
of hadrons at low temperatures.  First, the author calculates the meson profile function at low temperatures using the universal action derived in Ref \cite{15}. The corresponding EOM (Equation of motion) for the Fourier transform of the bulk-to-boundary profile function of mesons $\phi_{n}(r,T)$ in Euclidean metric is as follows: 
\begin{equation}
\left[-\frac{d^{2}}{dr^{2}}+U(r,T)\right]\phi_{n}(r,T)=M_{n}^{2}(T)\phi_{n}(r,T).
\label{16}
\end{equation}
 $U(r, T)$ is the effective potential and is consist of the  temperature-dependent and non-dependent parts:
\begin{equation}
U(r,T) =U(r) + \Delta U(r,T).
\label{17}
\end{equation}
Explicit forms of the $U(r)$ and $\Delta U(r,T)$ terms were given as
\begin{equation}
U(z)=k^{4}r^{2}+\frac{(4m^{2}-1)}{4r^{2}},
\label{18}
\end{equation}
\begin{equation}
\Delta U(r,T)=2\Delta(T)k^{4}r^{2}.
\label{19}
\end{equation}
 
 The meson mass spectrum $ M_{n}^{2}$ is shown the following form by the sum of zero and finite temperature parts:
\begin{equation}
 M_{n}^{2}(T) =\ M_{n}^{2}(0)+\Delta M_{n}^{2}(T),
 \label{20}
\end{equation}
\begin{equation}
\Delta M_{n}^{2}(T)=\rho(T)M_{n}^{2}(0) + \frac{R\pi^{4}T^{4}}{k^{2}},
\label{21}
\end{equation}
\begin{equation}
M_{n}^{2}(0)=4k^{2}\left(n+\frac{m+1}{2}\right), 
\end{equation}\\ 
$R =(6n-1)(m+1).$
\label{22}
In the low-temperature case, the hadronic mass spectrum is:
\begin{equation}
 M_{n}^{2}(T) =\ M_{n}^{2}(0)+\Delta M_{n}^{2}(T),
\end{equation}
\begin{equation}
\Delta M_{n}^{2}(T)=\rho_{T}M_{n}^{2}(0) + \frac{R\pi^{4}T^{4}}{k^{2}},
\end{equation}
\begin{equation}
R =(6n-1)(m+1).
\end{equation}
 The  profile  function of mesons in general form is given \cite{15} as follows:
  \begin{equation}
 \phi_{n}(r,T)=\sqrt{\frac{2\Gamma(n+1)}{\Gamma(n+m+1)}}K(T)^{m+1}r^{m+\frac{1}{2}}e^{-\frac{K(T)^{2}r^{2}}{2}}L_{n}^{m}(K(T)^{2}r^{2}).
\end{equation}
 For the meson with two quarks $m=L$. By taking m=1 for the vector and axial vector mesons, we have replaced $\phi_{n}(r,T)$ by $M_{0}(r,T)$ as the meson profile function in the expression of the coupling constant.
\section{RARITA-SCHWINGER FIELDS  AT FINITE TEMPERATURE}

According to the principle of holographic duality, the fields set against the $\Delta$ baryon operators and the nucleon operator with a defined spin 1/2 at the boundary of the AdS space-time differ from each other. Thus, while the Dirac field corresponds to nucleons with spin 1/2, the $\Delta$ baryon operators with spin 3/2 are opposed to  the Rarita-Schwinger fields $\Psi_{M}$  within this space-time.
\cite{20, 21, 22, 23}.
 The action for Rarita-Schwinger field at finite temperature can be written correspond to zero temperature case as follows:
\begin{equation}\label{RS}
\int
d^5x\sqrt{G}\left(i\bar{\Psi}_A\Gamma^{ABC}D_B\Psi_{C}-m_1\bar{\Psi}_A\Psi^A-m_2\bar{\Psi}_A\Gamma^{AB}\Psi_B\right)\,,
\end{equation}
where $\Psi_{A}=e_{A}^{M}\Psi_{M}$ and we used notations
$\Gamma^{ABC}=\frac{1}{3!}\Sigma_{\rm{perm}}(-1)^p\Gamma^A\Gamma^B\Gamma^C=\frac{1}{2}(\Gamma^B\Gamma^C\Gamma^A
-\Gamma^A\Gamma^C\Gamma^B)$ and
$\Gamma^{AB}=\frac{1}{2}[\Gamma^A\,,\Gamma^B]$. The Rarita-Schwinger
equations in $AdS_5$  are  written as
\begin{equation}
i\Gamma^A\Big(D_A \Psi_B - D_B \Psi_A\Big) - m_{-} \Psi_B + \frac
{m_{+}}{3}\Gamma_B\Gamma^A\Psi_A = 0
\end{equation}
where $m_\pm = m_1 \pm
m_2$. $m_1$ and $m_2$  correspond to those of spinor
harmonics on $S^5$ of $\text{AdS}_5 \times S^5$ \cite{24}.

As is well known, the Rarita–Schwinger field contains the states of the field with spin 1/2 in addition to spin 3/2 components. In 4-dimensional space, the components with spin 1/2 are eliminated by imposing the Lorentz condition on this field:
\begin{equation}
\gamma^{\mu}\Psi_{\mu}=0\,.
\end{equation}
The following Lorentz-covariant constraint will project out one of the spin-1/2 components from the Rarita-Schwinger fields in five dimensional space correspond to four dimensional space.
\begin{equation}
e_{A}^{M}\Gamma^{A}\Psi_{M}=0\,,
\end{equation}
which then gives $\partial^{M}\Psi_{M}=0$ for a free particle if combined with equations of motion.

This field has extra spin-1/2, $\Psi_{z}$, if decreased to four dimensional space-time at finite temperature.
 By
choosing  $\Psi_r = 0$, it is possible to reduce the extra spin-1/2 degrees of freedom, because there is no extra spinor at finite temperature.
\begin{equation}
(iz\Gamma^{A}\Psi_{A} +2i\Gamma^{5} - m_{-}) \Psi_{\mu}= 0,
\end{equation}
$\quad(\mu=0,1,2,3)$
\begin{equation}
\Psi_{M(R)}=\frac{1}{2}(1+\gamma^{5})\Psi_{NE}
\end{equation}
\begin{equation}
\Psi_{M(L)}=\frac{1}{2}(1-\gamma ^{5})\Psi_{M}
\end{equation}
\begin{equation}
[\partial _{r}^{2}-\frac{2(m_{-}+K(T)^{2}r^{2})}{r}\partial _{r}+\frac{2(m_{-}-K(T)^{2}r^{2})}{r^{2}}+p^{2}]\Psi_{L}=0
\end{equation}
\begin{equation}
[\partial _{r}^{2}-\frac{2(m_{-}+K(T)^{2}r^{2})}{r}\partial _{r}+p^{2}]\Psi_{R}=0
\end{equation}
The solution of these equations with the polynomial is similar to the fermion field, as follows:
\begin{eqnarray}
f_{n}^{L}(r,T)=\sqrt{\frac{2\Gamma (n+1)}{\Gamma (n+m_{L}+1)}}K^{m_{L}+1}r^{m_{L}+\frac{1}{2}}e^{-\frac{K^{2}r^{2}}{2}}L_{n}^{m_{L}}\left(K^{2}r^{2}\right), \nonumber\\
f_{n}^{R}(r,T)=\sqrt{\frac{2\Gamma (n+1)}{\Gamma (n+m_{R}+1)}}K^{m_{R}+1}r^{m_{R}+\frac{1}{2}}e^{-\frac{K^{2}r^{2}}{2}}L_{n}^{m_{R}}\left(K^{2}r^{2}\right),
\label{43}
\end{eqnarray}
 where
$m_{L,R}=m\pm\frac{1}{2}$.
To describe the spin 3/2 baryons ($\Delta$ resonances) in the soft-wall model at finite temperature 
one has to introduce a pair of Rarita-Schwinger fields in the
bulk, $\Psi_1^A$ (for the left-handed spin-3/2) and $\Psi_2^A$ (for the right-handed spin-3/2), which obey the above
Rarita-Schwinger equations. The author has considered Ref \cite{14} to describe the spin 1/2 baryons (nucleons) profile functions in the soft-wall model at a finite temperature which obeys the Dirac equation in the
bulk \cite {16}.  The left-handed spin-1/2 baryon profile function is given by  $F_1^A (r, T)$, whenever the right-handed spin-1/2 baryon is given by $F_2^A(r, T)$. Note that the profile functions of spin 1/2 and 3/2 baryon are the same in a certain approximation. 
The profile functions of $\Delta$ baryon $f_{m}(r,T)$ and nucleon $F_{n}(r,T)$ obey normalization conditions as 
\begin{equation}
\int_{0}^{\infty}dr e^{-\frac{3}{2}A(r)}f_{m}^{L,R}(r,T)F_{n}^{L,R}(r,T)=\delta_{mn} 
\label{44} 
\end{equation} 
\section{STRONG COUPLING CONSTANTS $g_{\rho \Delta \Delta}(T)$, $g_{a_{1} \Delta \Delta}(T)$, AND $g_{\rho N \Delta}(T)$ AT FINITE TEMPERATURE}
In this section the author interested in  the properties of $\Delta$ resonances in AdS/QCD at finite temperature.
The resonance to nucleons of $\Delta$ baryons are very necessary in the study  of  nucleon potential, as the decay channel of $\Delta$ resonance at finite temperature. Generally, the interaction Lagrangian is constructed based on the gauge invariant of the model and contains a term of minimal gauge interaction of the meson field with the fermions corresponding zero temperature limit \cite{18}. 
Meson-baryon transition coupling constants have been obtained from the action by including the thermal dilaton field in the action. 

 The thermal couplings have  expressions of thermal profile functions of the bulk fields. It has calculated the terms of thermal action in the momentum space and taken the variation derivative from Lagrangian terms. This variation gives us the following contribution of each Lagrangian term to the nucleon current:

\begin{equation}
{\cal L}_{\pi \Delta \Delta} =
\bar\Psi_1^{\mu} \Gamma^r A_r \Psi_{1\mu} -\bar\Psi_2^{\mu} \Gamma^r
A_r \Psi_{2\mu}
\end{equation}
here $\Gamma$ is Dirac matrices in the reference frame.  $\Gamma^{r}$ matrices are chosen as  $ \Gamma^{r}=(\gamma^{\mu },\ -i\gamma ^{5}) $.
The expression for the five dimensional spinors has following form:
$\Psi^\sigma _{L,R}(p,r) = \sum_n F^{(n)}_{L,R}(p,r) \psi^{(n)\sigma}_{L,R
} (p)$, 
The 4D
coupling constant for pions as
\begin{eqnarray}
g^{(0)nm}_{\pi \Delta \Delta}(T) &=& \! -\int^{\infty}_0 \frac{dr}{2r^2}
\left[M_{0}(r,T)\Big(f^{(n)*}_{1L}(r,T) f^{(m)}_{1R}(r,T) - f^{(n)*}_{2L}(r,T)
f^{(m)}_{2R}(r,T)\Big)\right.
\end{eqnarray}
The Lagrangian of $\rho$ meson is as follows \cite{25}:
\begin{equation}
{\cal L}_{\rho \Delta \Delta} =
\bar\Psi_1^{\nu} \Gamma^{\mu} V_{\mu} \Psi_{1\nu} +\bar\Psi_2^{\nu} \Gamma^{\nu}
V_{\nu} \Psi_{2\nu}
\end{equation}
For the 5D spinors
$\Psi^\sigma _{L,R}(p,r) = \sum_n F^{(n)}_{L,R}(p,r) \psi^{(n)\sigma}_{L,R
} (p)$, the 4D
coupling constant for $\rho$ meson is as follows:
\begin{eqnarray}
g^{(0)nm}_{\rho \Delta \Delta}(T) &=& \! -\int^{\infty}_0 \frac{dz}{r^2}
M_{0}(r,T)\left[\Big(f^{(n)*}_{1L}(r,T) f^{(m)}_{1L}(r,T) + f^{(n)*}_{2L}(r,T)
f^{(m)}_{2L}(r,T)\Big)\right.
\end{eqnarray}

 $a_{1}$ meson coupling constant arise from the bulk gauge
coupling constant \cite{26} and the minimal interaction Lagrangian terms has following form:
\begin{equation}
{\cal L}_{a_{1}\Delta\Delta}=\bar\Psi_{1}^{M}e_{A}^{M}\Gamma^{M}
A_{M}\Psi_{1M}-\bar\Psi_{2}^{M}e_{A}^{M}\Gamma^{M}A_{M}\Psi_{2M},
\end{equation}
the magnetic gauge interaction or Pauli terms ${\cal L}_{a_{1} \Delta \Delta}$ is as follows:
\begin{equation}
{\cal L}_{a_{1} \Delta \Delta} = k_{1}\bar\Psi_1^{M}e_{A}e_{B}^{N}\Gamma^{MN}
F_{MN}\Psi_{1M} -\bar\Psi_2^{M}e_{A}^{M}e_{B}^{N}\Gamma^{MN} F_{MN}
\Psi_{2M},
\end{equation}
Thus, after making certain simplifications, we obtain from the minimum interaction Lagrangian terms that $a_{1}$ meson $\Delta$ baryon minimal coupling constant  $g^{(0)nm}_{a_{1} \Delta \Delta}(T)$ in the framework of the soft-wall framework 
which can be written as
\begin{equation}
g^{(0)nm}_{a_{1} \Delta \Delta}(T) = \int^{\infty}_0 \frac{dr}{r^2} M_{0}(r, T)
\Big(f^{(n)}_{1R}(r,T)^{2}-f^{(m)}_{1L}(r,T)^{2}\Big).
\end{equation}
 $M_{0}(r, T)$ is expression of the
profile function of a meson in the ground state.
The additional contributions to $g^{(1)nm}_{a_{1} \Delta \Delta}(T)$ coupling constant can arise from the  magnetic
type of interaction in the bulk of AdS space-time as following form:
\begin{equation}
g^{(1)nm}_{a_{1} \Delta \Delta}(T) = \frac{k_{1}}{2}\int^{\infty}_0 \frac{dr}{r^2} M_{0}^{`}(r,T)
\Big(f^{(n)}_{1L}(r,T)^{2} + f^{(n)}_{1L}(r,T)^{2} )\Big).
\end{equation}
The $\pi$ meson transition coupling constant have been obtained from the gauge-
invariant coupling constant of gauge fields with  $\Delta$ resonances and nucleons
 at finite temperature. The Lagrangian for this fields is given by
\begin{eqnarray}\label{Sfnd}
{\cal
L}_{FN\Delta}&=&\left[\alpha_1\Big(\bar{\Psi}_1^M\Gamma^N(F_L)_{MN}N_1-(1\leftrightarrow
2 \,\, \&\,\, L\leftrightarrow R)\Big)\right. \nonumber\\
\end{eqnarray}
where $\alpha_{1}$ is a parameter \cite{18}. The terms
 contribute to the 4D $\rho$ meson thermal coupling constant.By KK reduction of 5D spinors as $\Psi_{iL,R}(p,z) =
\sum_n f^{(n)}_{iL,R}(p,z) \psi^{(n)}_{L,R} (p)$  for nucleons and
$\Psi_{L,R}(p,z) = \sum_n F^{(n)}_{L,R}(p,z) \psi^{(n)}_{L,R} (p)$ 
for $\Delta$ resonances, one can write the pion-nucleon-$\Delta$
couplings as,
\begin{eqnarray}
g^{nm}_{\pi N \Delta}(T) &=& -f_{\pi}\int^{\infty}_0 dz
\left[\frac{M_{0}(r, T)}{r^{2}}\Big(\kappa\big(F^{(n)*}_{1L}(r,T) f^{(m)}_{1R}(r,T) +
F^{(n)*}_{2L}(r,T) f^{(m)}_{2R}(r,T)\big)\right]
\end{eqnarray}
and similarly the rho-nucleon-$\Delta$ couplings at finite temperature as,
\begin{eqnarray}
g^{nm}_{\rho N \Delta}(T) &=& \int^{\infty}_0 dr
\left[\frac{M_{0}(r, T)}{r^2}\Big(\kappa \big(F^{(n)*}_{1L}(r,T)
f^{(m)}_{1R}(r,T) - F^{(n)*}_{2L}(r,T) f^{(m)}_{2R}(r,T)\big)\right]
\end{eqnarray}
\section{Numerical Results}
The coupling constants at finite temperature  $g_{\rho \Delta \Delta}(T)$, $g_{a_{1} \Delta \Delta}$(T), and $g_{\rho N \Delta}(T)$  consists in numerical calculation of the integrals, and in numerically plotting their temperature dependencies by means of the  Mathematical package. The author presents  numerical results for the parameters $N_{f}=2$, $F=87$ MeV, $N_{f}=3$, $F=100$ MeV, $N_{f}=4$, $F=130$ MeV and flovour $N_{f}=5$, $F=140$ MeV. These sets of parameters were taken  from \cite{18}
There are free parameters $k$, $k_{1}$, $m_{q}$ and $\Sigma$ in this work. The value of parameter $k=383$ MeV \cite{15} and parameter $\overline{k}=-733$ \cite{18}. 
The parameters  $k_{1}$  are fixed at the values  $k_{1}=-0.78$ $ GeV^{3}$ in the \cite{26}. The $\Sigma=0.368^{3}$ $MeV^{3}$ value and the $m_{q}=0.0023$ $GeV$ value of these parameters were found from the fitting of the $\pi$ meson mass \cite{27}. 
Having an idea of the relative contributions of different flavors of hadrons, the author present results for the temperature dependencies of the various numbers separately. 

  In Fig. 1 the $\rho$ meson $\Delta$ baryon coupling constant $g^{0}_{\rho \Delta \Delta}(T)$,  In Fig 2 the $a_{1}$ meson $\Delta$ baryon coupling constant $g^{0}_{a_{1} \Delta\Delta}(T)$ and In Fig. 4 the $\rho$ meson-nucleon-$\Delta$ -transition coupling constant $g_{\rho N \Delta}(T)$ were plotted at the parameter $N_{f}=2$, $F=87$ MEV, $N_{f}=3$, $F=100$ MEV, $N_{f}=4$, $F=130$ MEV, and $N_{f}=5$, $F=140$ MEV. 
  In Fig 3 the author also has considered these dependencies for $g^{1}_{a_{1} \Delta \Delta}(T)$ of the nucleons and drawing graphs for the different number of flavors. The purple graph curve represents two $N_{f}=2$, $F=87$ $MeV$, green  one curve shows the  three $N_{f}=3$, $F=100$ MeV, orange curve shows the  four  $N_{f}=4$, $F=130$ MeV , and blue one shows the five flavor $N_{f}=5$, $F=140$ MeV of thermal  minimal coupling constants in the figures. 
  This analysis shows a weak dependence on the parameter $N_{f}$ of the coupling constants. 
\newpage
\section{SUMMARY}
 In the present work, the author study the temperature dependency of minimal coupling constants of  meson-$\Delta$ baryon and meson-nucleon to $\Delta$ baryon transition 
in the framework of the soft-wall model of holographic QCD. It has been observed that the value of all terms become zero at the same point near the Hawking temperature by increasing of temperature. The minimal coupling constants of  meson-$\Delta$ baryon and meson-nucleon to $\Delta$ baryon transition may be of use in the deeper study of the nucleon-delta baryon transition and in understanding processes of the early Universe.

\begin{figure}[!ht]
\centering
\includegraphics[scale=0.6]{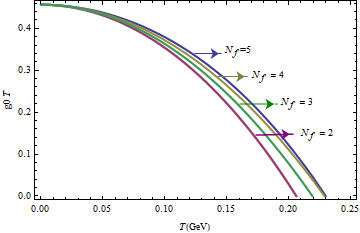}
\caption{The temperature dependence of $g^{0}_{\rho \Delta \Delta}(T)$  for $N_{f}=2$, $F=87$ MEV, $N_{f}=3$, $F=100$ MEV, $N_{f}=4$, $F=130$ MEV, $N_{f}=5$, $F=140$ MEV.}
\end{figure}
\begin{figure}[!ht]
\centering
\includegraphics[scale=0.6]{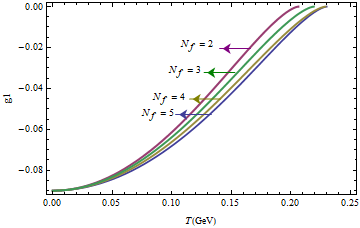}
\caption{The temperature dependence of $g^{0}_{a_{1} \Delta\Delta}(T)$  for $N_{f}=2$, $F=87$ MEV, $N_{f}=3$, $F=100$ MEV, $N_{f}=4$, $F=130$ MEV, $N_{f}=5$, $F=140$ MEV.}
\end{figure}
\begin{figure}[!ht]
\centering
\includegraphics[scale=0.6]{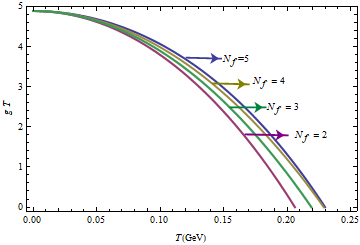}
\caption{The temperature dependence of $g^{1}_{a_{1} \Delta  \Delta}(T)$  for $N_{f}=2$, $F=87$ MEV, $N_{f}=3$, $F=100$ MEV, $N_{f}=4$, $F=130$ MEV, $N_{f}=5$, $F=140$ MEV.} 
\end{figure}
\begin{figure}[!ht]
\centering
\includegraphics[scale=0.6]{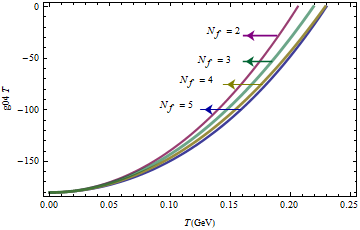}
\caption{The temperature dependence of $g_{\rho N \Delta}(T)$  for $N_{f}=2$, $F=87$ MEV, $N_{f}=3$, $F=100$ MEV, $N_{f}=4$, $F=130$ MEV, $N_{f}=5$, $F=140 $ MEV.}
\label{fig:Figure 5}
\end{figure}
\end{document}